\documentclass[aps,twocolumn,nofootinbib,prd,UTF8]{revtex4-1}
\usepackage{epsfig}
\usepackage[colorlinks,linkcolor=blue,anchorcolor=blue,citecolor=blue,urlcolor=blue,breaklinks=true]{hyperref}
\usepackage{graphics}
\usepackage{color}
\usepackage{amsmath}
\usepackage{float}
\usepackage{tabularx}
\usepackage{url}
\begin{document}
\bibliographystyle{unsrt}

\author{Song-Ju Lei$,^{1,}$}~\email[]{Email: leisongju@smail.nju.edu.cn},\author{Run-Lin Liu$^{1}$}
\address{$^{1}$School of Physics, Nanjing University, Nanjing 210093, China}

\title{Nonlinear susceptibilities of heavy ion collisions within the Polyakov-loop-extended Nambu-Jona-Lasinio model }
\begin{abstract}
The baryon-number susceptibilities is correlated to the fluctuations obtained in experiments, we can theoretically calculate the susceptibility and compare it with the experimental fluctuations data. In this paper, we calculate the baryon-number susceptibilities from the Polyakov-loop-extended Nambu-Jona-Lasinio (PNJL) model, then compare them with the the heavy ion collision experimental fluctuation data and the results from other models, lattice QCD and the Dyson-Schwinger equations (DSEs) approach. Our results are in line with the experiment and very similar to the results of DSEs at experimental points, which shows that the PNJL model is suitable for studying this issue.\end{abstract}


\maketitle

\section{INTRODUCTION}
In recent decades, the quantum chromodynamics (QCD) phase transition and corresponding phase diagram have been studied extensively through a series of theoretical tools and heavy ion collision experiments~\cite{adams2005experimental}. In particular, QCD transitions from hadronic matter to the quark-gluon plasma  (QGP) is considered occur in the heavy ion collision experiments~\cite{asakawa2000fluctuation}. This process is a smooth crossover at small baryon chemical potential and high temperature, and a first-order phase transition is expected at high baryon chemical region. The intersection between first-order phase transition and crossover region is called QCD critical end point (CEP), people want to find CEP by experiment. 

From the usual view, quark-number (or baryon-number) susceptibility (the second order) will show some singularity~\cite{stephanov1998signatures,stephanov1999event} near the CEP. Many phenomenological models~\cite{hatta2003universality,costa2008thermodynamics,sasaki2007susceptibilities,schaefer2007susceptibilities,redlich2008density,fukushima2008phase,cui2013wigner,jiang2013chiral,Cui2013aba,Wang2015tia,Shi2015ufa,Cui2015xta,Cui2017ilj} have been used and lattice QCD~\cite{allton2005thermodynamics,gavai2005critical,philipsen2007lattice} calculations are used to find  the location of the CEP.  Baryon number fluctuations have long been considered to be closely 
related to phase transitions, experiments such as STAR's beam energy scan (BES) program and
PHENIX experiments at the Relativistic Heavy-Ion Collider (RHIC)~\cite{aggarwal2010higher} have been carried out to measure them. In our work, we used the experimental data of RHIC, which obtained the fluctuation of baryon number in Au + Au collisions. The $n$th cumulant of baryon-number
fluctuations is proportional to the $n$th order of baryon-number susceptibilities~\cite{gupta2011scale,asakawa2000fluctuation,jeon2000charged}, so we need to calculate the nth order of baryon-number susceptibilities theoretically in order to compare them with the experimental values. 

It can be found that the quark-number density determined by the corresponding dressed quark propagator at finite chemical potential~\cite{zong2008calculation}. There are variational methods that one can use a bare propagator and obtain nonperturbative results, such as hard thermal loop perturbation theory, optimized perturbation theory, and so on. However, we don't use variational methods. Here, we will get the dressed quark propagator through the PNJL model~\cite{R2006Polyakov}, which is the Nambu-Jona-Lasinio (NJL) model improved with the Polyakov loop. In recent years,  the PNJL model has been successfully used to describe the thermodynamics of QCD with two and two-plus-one flavors, it allows for a simultaneous computation of quantities sensible to confinement
and chiral symmetry breaking~\cite{Ruggieri2011The}. Some other models such as NJL model, the chiral random matrix model, the linear sigma model and chiral perturbation theory are based on chiral symmetry, but they all lack any dynamics coming form the Polyakov loop.
~\cite{fukushima2004chiral}.

The rest of this paper is organized as follows: In Sec.\ref{sec2} nonlinear susceptibilities and a mean field description of the PNJL model is presented. In Sec.\ref{sec3} we present our calculation results, which were transformed into experimentally observable results for comparisons and research. Finally, in Sec.\ref{sec4} we will summarize our results and give the conclusions.
\section{PNJL MODEL AND EFFECTIVE POTENTIAL}\label{sec2}
The renormalized partition function of QCD at zero temperature and finite chemical is of the form~\cite{zong2008calculation}

\begin{equation}
\begin{aligned}
\mathcal{Z}[\mu]=&\int \mathcal{D} \overline{q}_{R} \mathcal{D} q_{R} \mathcal{D} A_{R} \exp \{-S_{R}\left[\overline{q}_{R}, q_{R}, A_{R}\right]+  \\
&\int d^{4} x \mu Z_{2} \overline{q}_{R}(x) \gamma_{4} q_{R}(x)\}
\end{aligned}
\end{equation}
where $S_{R}\left[\overline{q}_{R}, q_{R}, A_{R}\right]$ is standard renormalized Euclidean QCD action, $q_{R}$ is
the renormalized quark field with three flavors and three colors; $Z_{2}=Z_{2}\left(\zeta^{2}, \Lambda^{2}\right)$ is quark wave-function renormalization constant, where $\zeta$
is the renormalization point and $\Lambda$ is the regularization mass-scale. By leave the ghost field term and its integration measure
to be understood, the pressure density $\mathcal{P}(\mu)$ is given by
\begin{equation}
\mathcal{P}(\mu)=\frac{1}{\mathcal{V}} \ln \mathcal{Z}[\mu]
\end{equation}
where $\frac{1}{\mathcal{V}}$ is the four-volume normalizing factor. And the quark-number density is
\begin{equation}
\rho(\mu)=\frac{\partial \mathcal{P}(\mu)}{\partial \mu}
\end{equation}
From Eq.(1), (2) and (3), we can easily obtain the result~\cite{Zong2008sm}
\begin{equation}
\rho(\mu)=(-) N_{c} N_{f} Z_{2} \int \frac{d^{4} p}{(2 \pi)^{4}} \operatorname{tr}\left[G[\mu]({p}) \gamma_{4}\right] \label{pythagorean}
\end{equation}

Where $N_{c}$ and $N_{f}$ are, respectively, the number of colors and flavors; $G[\mu]({p})$ is the quark propagator; From this formula, we can see that in the case of zero temperature and limited chemical potential, the quark number
density is determined only by the dressed quark propagator at finite chemical potential,
by ignore the $\mu$ dependence
of the dressed gluon propagator and assume that the dressed quark propagator at finite $\mu$ is analytic in the neighborhood
of $\mu$=0, then we can obtain the following expression~\cite{hou2005new}
\begin{equation}
G[\mu]({p})^{-1}=G(\tilde{p})^{-1}
\end{equation}
Where $G[\mu]({p})^{-1}=i\gamma$ · ${p}+M$, $M$ is  effective quark mass; $\tilde{p}=\left(\vec{p}, p_{4}+i \mu\right)$. By some mathematical methods, like the matsubara frequency, we can get the expression
 for the quark-number density at finite temperature
\begin{equation}
\rho(T, \mu)=(-) N_{c} N_{f} T \sum_{i=-\infty}^{+\infty} \int \frac{d^{3} \vec{p}}{(2 \pi)^{3}} \operatorname{tr}\left[G\left(\tilde{p}_{n}\right) \gamma_{4}\right]
\end{equation}
the fourth component of momentum is $\omega_{n}+i\mu$, and the fermion frequencies $\omega_{n}=(2n+1)\pi T$.  A baryon consists of three quarks, so the quark-number density is three times as many as the baryon-number density. Then the $\rho_{B}$ to the baryon chemical potential $\mu_{b}$'s (n-1) order derivatives
are defined by the nonlinear susceptibilities of baryons of order n~\cite{gavai2003pressure}.
\begin{equation}
\begin{aligned} \chi_{B}^{(n)} &=\frac{\partial^{n-1}}{\partial \mu_{B}^{n-1}} \rho_{B}=\frac{\partial^{n-1}}{3^{n} \partial \mu^{n-1}} \rho(T, \mu) \\ &=(-) \frac{N_{c} N_{f} T}{3^{n}} \sum_{i=-\infty}^{+\infty} \int \frac{d^{3} \vec{p}}{(2 \pi)^{3}} \operatorname{tr}_{\gamma}\left[\frac{\partial^{n-1} G\left(\tilde{p}_{n}\right)}{\partial \mu^{n-1}} \gamma_{4}\right] \end{aligned}
\end{equation}
And then by taking the trace and summing the frequencies, we can get
\begin{equation}
\begin{aligned} &\rho(T, \mu)= \\ & (-) \frac{N_{c} N_{f} T}{\pi} \int (\frac{1}{e^{\frac{\sqrt{M^2+\vec{p}^2} - \mu}{T}} + 1} - \frac{1}{e^{\frac{\sqrt{M^2+\vec{p}^2} + \mu}{T}} + 1}) p^2 d p\end{aligned}
\end{equation}
So what we're going to do is figure out the effective quark mass $M$ in the PNJL model.

The Lagrangian density of two flavors of equal-mass quarks in the PNJL model is
\begin{equation}
\begin{aligned} \mathcal{L}=& \overline{q}(\gamma \cdot D+m) q-G\left[(\overline{q} q)^{2}+\left(\overline{q} i \gamma_{5} \tau q\right)^{2}\right] \\ &+\mathcal{U}(\Phi, \overline{\Phi} ; T) \end{aligned}
\end{equation}
 where $m$ is the common current-quark mass; $D_{\mu} = {\partial_\mu}+i A_{\mu}$, with $A_{\mu}(x)=g_{s} A_{\mu}^{a} \lambda^{a} / 2$ describe the matrix-valued gluon field configuration that fits the model; G is the four-fermion interaction strength and $\mathcal{U}$ is a Polyakov-loop effective potential.

Adopting the mean-field approximation, we take $L=\overline{L}$ from the beginning as in~\cite{Ratti2006wg}, the effective potential of the model can be separately expressed as~\cite{cui2016critical,Pan2016ecs},
\begin{equation}
\Phi=\frac{1}{N_{c}} \operatorname{Tr}_{c} L=\overline{\Phi}
\end{equation}
according to the classical background field in Eq.(11), the effective potential of Polyakov rings is expressed as follows~\cite{roessner2007polyakov}
\begin{equation}
\begin{aligned} \frac{1}{T^{4}} \mathcal{U}(\overline{\Phi}, \Phi ; T)=& \frac{1}{T^{4}} \mathcal{U}(\Phi ; T) \\=&-\frac{1}{2} a(T) \Phi^{2}+b(T) \\ & \times \ln \left[1-6 \Phi^{2}+8 \Phi^{3}-3 \Phi^{4}\right] \\ \end{aligned}
\end{equation}
with $(\overline{t}=T_{0} / T)$
\begin{equation}
a(\overline{t})=a_{0}+a_{1} \overline{t}+a_{2} \overline{t}^{2}, \quad b(\overline{t})=b_{3} \overline{t}
\end{equation}
For the sake of reproduce the lattice results for pure-gauge QCD chromodynamics and the T-dependence of Polyakov loop, we set these parameters and list them in Table {I}~\cite{roessner2007polyakov}.
\begin{table}
\renewcommand\arraystretch{1.8}
\centering
\begin{tabular}{p{0.5cm}<{\centering}p{3cm}<{\centering}p{1cm}<{\centering}p{3cm}<{\centering}p{0.5cm}<{\centering}}
\toprule
$a_0$&$a_1$&$a_2$&$b_3$&$T_0$\\
\hline
$3.51$&$-2.47$&$15.2$&$-1.75$&$190$\\
\hline
\hline
$m$&$\Lambda$&$G\Lambda^2$&$\alpha_1$&$\alpha_2$\\
\hline
$5.5$&$631.5$&$2.2$&$0.2$&$0.2$\\
\toprule
\end{tabular}
\caption{  Relevant parameters of the PNJL model ,with dimensioned quantities in MeV.}
\end{table}
The four-fermion coupling, $G$, is considered a constant~\cite{fukushima2004chiral},
The effective potential of the PNJL model is obtained by means of the mean-field approximation~\cite{Rossner2007ik,Ruggieri:2011xc},
\begin{equation}
\begin{aligned} \Omega=& \Omega\left(M, \Phi ; T, \mu\right) \\=& \mathcal{U}(\Phi ; T)+\frac{(M-m)^{2}}{4 G}-4 N_{c} \int \frac{d^{3} \vec{p}}{(2 \pi)^{3}} \omega \\ &-8T \int \frac{d^{3} \vec{p}}{(2 \pi)^{3}} \ln \left[\mathcal{F}\right] \end{aligned}
\end{equation}
 where $M$ and $\mathcal{F}$ is
 \begin{equation}
 \omega=\sqrt{M^2+\vec{p}^2}
 \end{equation}
 \begin{equation}
\mathcal{F}_{\pm}=1+3 \Phi\left[\mathrm{e}^{-\frac{\omega_{\pm}}{T}}+\mathrm{e}^{-2 \frac{ \omega_{\pm}}{T}}\right]+\mathrm{e}^{-3 \frac{ \omega_{\pm}}{T}}
 \end{equation}
$\omega_{\pm}=\omega \pm \mu$. It's worth noting that, we have not yet explicitly solved the regularization problem of the PNJL model. The last term in the second line of Eq. (13) is a divergent quantity, a regularization procedure must be introduced. For the reasons explained in~\cite{Pan2016ecs}, we impose a hard cutoff on both the integrals, although, the last one is not divergent. We employ a hard cutoff $\Lambda$, which is listed in table {I} . Using that value, and $m$ and $g$ listed, we obtained a good description of in-vacuum pion properties.
At this point, we can determine the quark mass gap evolution with intensive parameters by solving two external conditions simultaneously:
 \begin{equation}
 \frac{\partial \Omega}{\partial M}=0=\frac{\partial \Omega}{\partial \Phi}
 \end{equation}
 We solved the above equation by iterative method, then obtained the high-order susceptibility by numerical differentiation method.

\section{RESULT}\label{sec3}
\begin{figure}[h]
\centering
\includegraphics[width=8.7cm,]{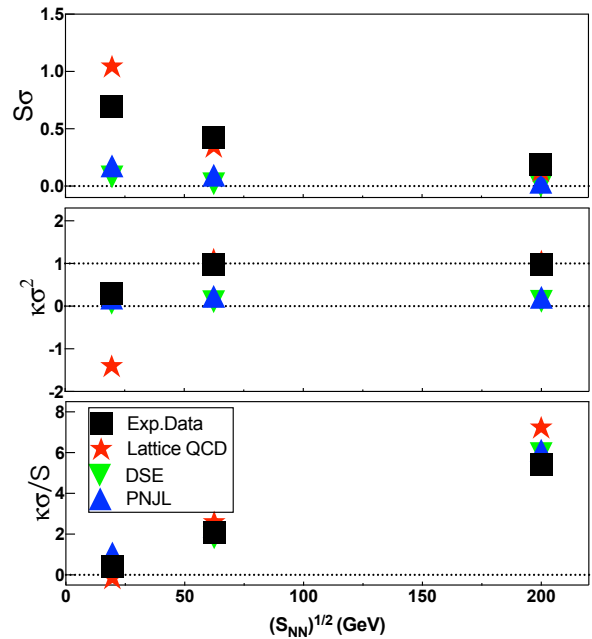}
\caption{Compare $S\sigma$, $\kappa\sigma^2$, and $\frac{\kappa\sigma}{S}$ of the the PNJL model results, lattice QCD, and experimental data at $\sqrt{S_{NN}}$ = 19.6, 62.4, 200 GeV. The black boxes for the experimental data, the red circle for lattice QCD, the blue upper triangle for the PNJL model and the green lower triangle for DSEs.}
\end{figure}
In order to compare our results with the experimental date and other methods, we mainly calculate the following formulas~\cite{gupta2011scale}
\begin{equation}
\begin{aligned} S \sigma &=\frac{T \chi_{B}^{(3)}}{\chi_{B}^{(2)}} \\ \kappa \sigma^{2} &=\frac{T^{2} \chi_{B}^{(4)}}{\chi_{B}^{(2)}} \\ \frac{\kappa \sigma}{S} &=\frac{T \chi_{B}^{(4)}}{\chi_{B}^{(3)}} \end{aligned}
\end{equation}
where $\sigma^2$ is the variance, $S$ is the skewness, and $\kappa$ is the kurtosis. In order to better compare with other researchers' work, we show $S\sigma$, $\kappa\sigma^2$ and $\frac{\kappa \sigma}{S}$ as a function of $\sqrt{S_{NN}}$ for $Au + Au$ collisions at RHIC in Fig.1. The top column represents the corresponding freeze-out chemical potential $\mu_B$ to $\sqrt{S_{NN}}$. Here we use the more commonly used empirical relationship to show the correlations between $\sqrt{S_{NN}}$ and the bulk properties ($\mu_B$ and $T$) of chemical freeze-out~\cite{cleymans1998unified,braun2009colloquium,cleymans2006comparison,andronic2004ultrarelativistic}.

\begin{table}
\caption{Correlation between $\sqrt{S_{NN}}$, temperature, baryon, and quark chemical potential.}
\renewcommand\arraystretch{1.8}
\centering
\begin{tabular}{p{2cm}<{\centering}p{2cm}<{\centering}p{2cm}<{\centering}p{2cm}<{\centering}}
\toprule
$\sqrt{S_{NN}}$ (GeV)&$T$ (MeV)&$\mu_B$ (MeV)&$\mu$ (MeV)\\
\end{tabular}
\centering
\renewcommand\arraystretch{1.2}
\begin{tabular}{p{2cm}<{\centering}p{2cm}<{\centering}p{2cm}<{\centering}p{2cm}<{\centering}}
\hline
$19.6$&$159$&$229$&$76$\\
$62.4$&$165$&$82$&$27$\\
$200$&$166$&$27$&$9$\\
\toprule
\end{tabular}
\end{table}

\begin{equation}
\begin{aligned} T\left(\mu_{B}\right) &=a-b \mu_{B}^{2}-c \mu_{B}^{4} \\ \mu_{B}(\sqrt{S_{N N}}) &=\frac{d}{1+e \sqrt{S_{N N}}} \\ \mu_B &=3\mu\end{aligned}
\end{equation}
where $a=0.166\pm0.002$ GeV, $b=0.139\pm0.016$ $\text{GeV}^{-1}$, $c=0.053\pm0.021$ $\text{GeV}^{-3}$, $d=1.308\pm0.028$ GeV, and $e=0.273\pm0.008$ $\text{GeV}^{-1}$~\cite{cleymans2006comparison}. According to the above formula, we list the corresponding values of $T,\mu_B,\mu$, and $\sqrt{S_{NN}}$ in Table {II}.
\begin{figure}
\centering
\includegraphics[width=8.7cm]{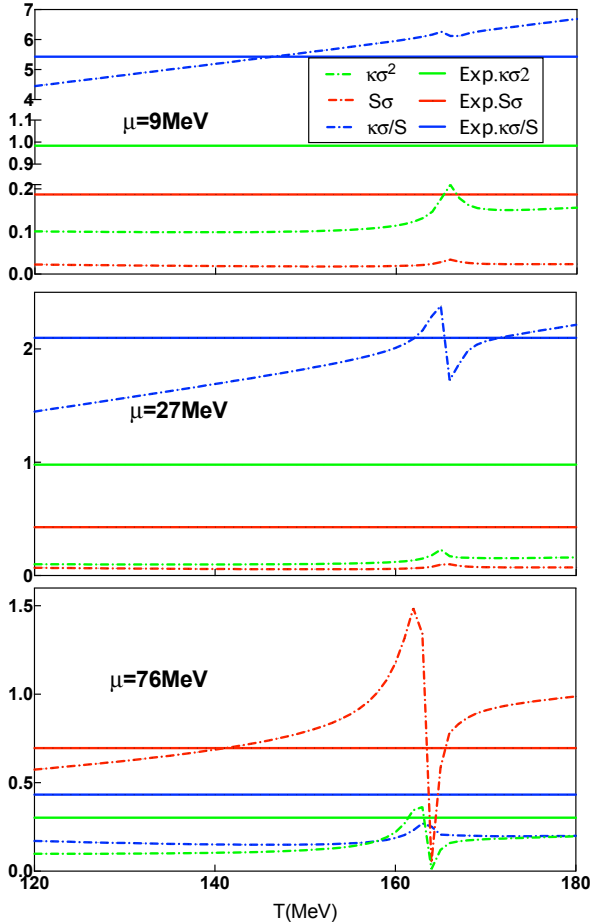}
\caption{$S\sigma$, $\kappa\sigma^2$, and $\frac{\kappa\sigma}{S}$ obtained by our calculation are respectively drawn as functions of $T$ and shown as curves in each plot. The corresponding experimental data of $S\sigma$, $\kappa\sigma^2$,  and $\frac{\kappa\sigma}{S}$ are drawn as straight lines and displayed in each figure.}
\end{figure}

The value of $\sqrt{S_{NN}}$ is for comparison with experimental data.The results that we obtained by the PNJL model are compared with lattice QCD~\cite{gavai2011lattice}, DSEs~\cite{zhao2014nonlinear} and experimental data~\cite{aggarwal2010higher}.

    As can be seen from Fig.1, the fitting of experiment of the PNJL model in $S\sigma$ and $\kappa\sigma^2$ is not as good as that of lattice QCD. However, we have a better fit at $\frac{\kappa\sigma}{S}$. This result is very similar to that of DSEs, so we used a similar approach to explore the reasons for the inconsistency with the experiment. We fixed the chemical potential $\mu$ at 9, 27 and 76 $MeV$, and we compute the curves of $S\sigma$, $\kappa\sigma^2$, and $\frac{\kappa\sigma}{S}$ as a function of temperature $T$, from 100 to 160 $MeV$.

    We present the curves of  $S\sigma$, $\kappa\sigma^2$,  and $\frac{\kappa\sigma}{S}$ obtained by our calculation as functions of  $T$ and $\mu$ in Fig.2 , where $\mu$ = 9, 27 and 76 $MeV$ respectively. We found that $S\sigma$ and $\kappa\sigma^2$ in our results were smaller than the experimental values, and $\frac{\kappa\sigma}{S}$ is larger than the experimental value. DSEs also encountered the same problem, fix $T$ at 160 and 166 MeV, $S\sigma$ and  $\kappa\sigma^2$ obtain by DSEs are all too small to compare with the experimental data when $\mu$ is less than 90 MeV. Therefore, we did the same calculation by the PNJL model, the results are shown in Fig.3. We find that there is an intersection between $S\sigma$ and the minimum experimental value, but it is smaller than other experimental values. And $\kappa\sigma^2$ is always smaller than experimental value no matter how $\mu$ changes. 
    
    Lowering the value of $m$, or $\Lambda$, or increasing the value of $G$ will bring our calculation closer to the experimental value. However, the parameters of the model are all obtained by fitting the meson mass $m_\pi$, the pion decay constant $f_\pi$ and the quark condensates $\langle\overline{q}q\rangle$~\cite{Lu2016uwy}, so we don't know how they should change with temperature and chemical potential. In this paper, we didn't fine-tune the parameters.
       
\begin{figure}
\centering
\includegraphics[width=9.4cm,]{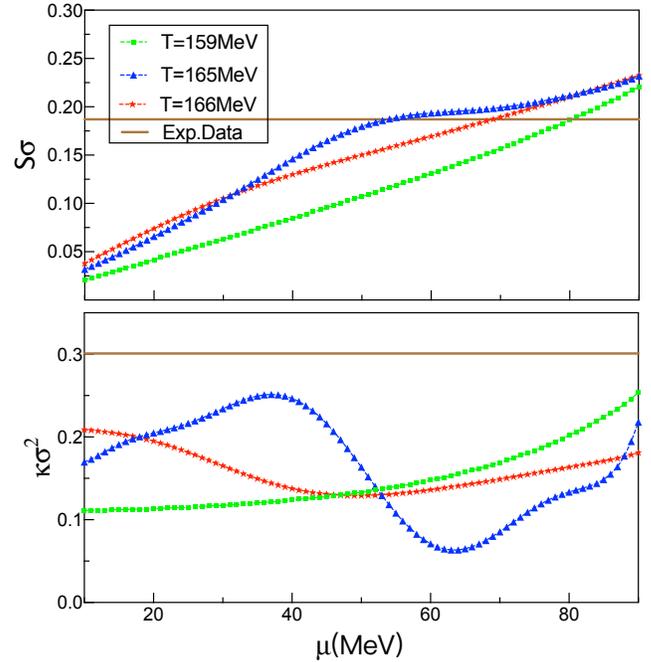}
\caption{Fix T at 159, 165 and 166 MeV,  $S\sigma$, $\kappa\sigma^2$ obtained by the PNJL model are  shown as a function of $\mu$ in each plot. We draw the smallest value of the corresponding experimental data at $\sqrt{S_{NN}}$ = 19.6, 62.4 and 200 GeV as a straight solid line and show it in the figure.}
\end{figure}

\section{SUMMARY}\label{sec4}
In this paper, we calculate baryon-number susceptibilities from the Polyakov-loop-extended Nambu-Jona-Lasinio models, then compare our results with the results from the heavy ion collision experiments and other models.

The curves of $S\sigma$, $\kappa\sigma^2$,  and $\frac{\kappa\sigma}{S}$ obtained by our calculation as functions of  $T$ and $\mu$  are very different from the DSEs curves in areas outside the scope of the experiment, but the results of the two different models were surprisingly consistent at several points in the experimental values. In addition, $S\sigma$ and  $\kappa\sigma^2$ of DSEs are smaller than the experimental value no matter how $\mu$ changes, and our results are similar to those of DSEs. This means that the different in DSEs and the PNJL model will lead to different peak point and CEP calculated by the two methods, and that is also why the curves of  $S\sigma$, $\kappa\sigma^2$,  and $\frac{\kappa\sigma}{S}$ as function of $T$ and $\mu$
 obtained by the two methods are different. The consistency of the two different methods at the experimental point shows the validity and the rationality of the two methods in dealing with the heavy ion collision experiments. However, $S\sigma$ and  $\kappa\sigma^2$ are smaller than the experimental results, which means that the PNJL model also has the same characteristics as DSEs, and there are still further reasons for us to explore. In general, the PNJL model is more concise and convenient than DSEs, but it still gives reasonable results. 
This proves that the PNJL model can be well used to deal with this issue. The difference between the results in Fig.2 and DSE also provides the correct direction for the experiment to verify the two models. Outside the existing experimental area, the two methods give significantly different results, so further research is needed.

Obviously, there's more that can be done here, such as our Polyakov loop potential is $\mu$-independent and we don't consider the impact of a vector interaction, $G_{V}(\overline{q}\gamma_{\mu}q)^2$, which may influence the results at high baryonic densities. There are some other work where Polyakov loop potenial is $\mu$-dependent~\cite{Ivanytskyi2019ojt} or $G_{V}(\overline{q}\gamma_{\mu}q)^2$~\cite{Buballa2003qv} is considered, our subsequent work will dive into these issues.
\section{ACKNOWLEDGMENTS}
The author would like to thank Dong Bai for helpful communications. This work is supported by the
 National Natural Science Foundation of China (Grant No. 11535004, 11761161001, 11375086, 11120101005, 11175085, 11235001 and 11975167).

\bibliographystyle{apsrev4-1}
\bibliography{bibfile}

\begin{thebibliography}{45}%
\makeatletter
\providecommand \@ifxundefined [1]{%
 \@ifx{#1\undefined}
}%
\providecommand \@ifnum [1]{%
 \ifnum #1\expandafter \@firstoftwo
 \else \expandafter \@secondoftwo
 \fi
}%
\providecommand \@ifx [1]{%
 \ifx #1\expandafter \@firstoftwo
 \else \expandafter \@secondoftwo
 \fi
}%
\providecommand \natexlab [1]{#1}%
\providecommand \enquote  [1]{``#1''}%
\providecommand \bibnamefont  [1]{#1}%
\providecommand \bibfnamefont [1]{#1}%
\providecommand \citenamefont [1]{#1}%
\providecommand \href@noop [0]{\@secondoftwo}%
\providecommand \href [0]{\begingroup \@sanitize@url \@href}%
\providecommand \@href[1]{\@@startlink{#1}\@@href}%
\providecommand \@@href[1]{\endgroup#1\@@endlink}%
\providecommand \@sanitize@url [0]{\catcode `\\12\catcode `\$12\catcode
  `\&12\catcode `\#12\catcode `\^12\catcode `\_12\catcode `\%12\relax}%
\providecommand \@@startlink[1]{}%
\providecommand \@@endlink[0]{}%
\providecommand \url  [0]{\begingroup\@sanitize@url \@url }%
\providecommand \@url [1]{\endgroup\@href {#1}{\urlprefix }}%
\providecommand \urlprefix  [0]{URL }%
\providecommand \Eprint [0]{\href }%
\providecommand \doibase [0]{http://dx.doi.org/}%
\providecommand \selectlanguage [0]{\@gobble}%
\providecommand \bibinfo  [0]{\@secondoftwo}%
\providecommand \bibfield  [0]{\@secondoftwo}%
\providecommand \translation [1]{[#1]}%
\providecommand \BibitemOpen [0]{}%
\providecommand \bibitemStop [0]{}%
\providecommand \bibitemNoStop [0]{.\EOS\space}%
\providecommand \EOS [0]{\spacefactor3000\relax}%
\providecommand \BibitemShut  [1]{\csname bibitem#1\endcsname}%
\let\auto@bib@innerbib\@empty
\bibitem [{\citenamefont {Adams}\ \emph {et~al.}(2005)\citenamefont {Adams}
  \emph {et~al.}}]{adams2005experimental}%
  \BibitemOpen
  \bibfield  {author} {\bibinfo {author} {\bibfnamefont {J.}~\bibnamefont
  {Adams}} \emph {et~al.} (\bibinfo {collaboration} {STAR}),\ }\href {\doibase
  10.1016/j.nuclphysa.2005.03.085} {\bibfield  {journal} {\bibinfo  {journal}
  {Nucl. Phys.}\ }\textbf {\bibinfo {volume} {A757}},\ \bibinfo {pages} {102}
  (\bibinfo {year} {2005})},\ \Eprint {http://arxiv.org/abs/nucl-ex/0501009}
  {arXiv:nucl-ex/0501009 [nucl-ex]} \BibitemShut {NoStop}%
\bibitem [{\citenamefont {Asakawa}\ \emph {et~al.}(2002)\citenamefont
  {Asakawa}, \citenamefont {Heinz},\ and\ \citenamefont
  {Muller}}]{asakawa2000fluctuation}%
  \BibitemOpen
  \bibfield  {author} {\bibinfo {author} {\bibfnamefont {M.}~\bibnamefont
  {Asakawa}}, \bibinfo {author} {\bibfnamefont {U.~W.}\ \bibnamefont {Heinz}},
  \ and\ \bibinfo {author} {\bibfnamefont {B.}~\bibnamefont {Muller}},\
  }\bibfield  {booktitle} {\emph {\bibinfo {booktitle} {{Quark matter 2001.
  Proceedings, 15th International Conference on Ultrarelativistic nucleus
  nucleus collisions, QM 2001, Stony Brook, USA, January 15-20, 2001}}},\
  }\href {\doibase 10.1016/S0375-9474(01)01418-X} {\bibfield  {journal}
  {\bibinfo  {journal} {Nucl. Phys.}\ }\textbf {\bibinfo {volume} {A698}},\
  \bibinfo {pages} {519} (\bibinfo {year} {2002})},\ \bibinfo {note}
  {[,398(2001)]},\ \Eprint {http://arxiv.org/abs/nucl-th/0106046}
  {arXiv:nucl-th/0106046 [nucl-th]} \BibitemShut {NoStop}%
\bibitem [{\citenamefont {Stephanov}\ \emph {et~al.}(1998)\citenamefont
  {Stephanov}, \citenamefont {Rajagopal},\ and\ \citenamefont
  {Shuryak}}]{stephanov1998signatures}%
  \BibitemOpen
  \bibfield  {author} {\bibinfo {author} {\bibfnamefont {M.~A.}\ \bibnamefont
  {Stephanov}}, \bibinfo {author} {\bibfnamefont {K.}~\bibnamefont
  {Rajagopal}}, \ and\ \bibinfo {author} {\bibfnamefont {E.~V.}\ \bibnamefont
  {Shuryak}},\ }\href {\doibase 10.1103/PhysRevLett.81.4816} {\bibfield
  {journal} {\bibinfo  {journal} {Phys. Rev. Lett.}\ }\textbf {\bibinfo
  {volume} {81}},\ \bibinfo {pages} {4816} (\bibinfo {year} {1998})},\ \Eprint
  {http://arxiv.org/abs/hep-ph/9806219} {arXiv:hep-ph/9806219 [hep-ph]}
  \BibitemShut {NoStop}%
\bibitem [{\citenamefont {Stephanov}(1999)}]{stephanov1999event}%
  \BibitemOpen
  \bibfield  {author} {\bibinfo {author} {\bibfnamefont {M.~A.}\ \bibnamefont
  {Stephanov}},\ }\bibfield  {booktitle} {\emph {\bibinfo {booktitle} {{Quark
  matter '99. Proceedings, 14th International Conference on ultrarelativistic
  nucleus nucleus collisions, QM'99, Torino, Italy, May 10-15, 1999}}},\ }\href
  {\doibase 10.1016/S0375-9474(99)85051-9} {\bibfield  {journal} {\bibinfo
  {journal} {Nucl. Phys.}\ }\textbf {\bibinfo {volume} {A661}},\ \bibinfo
  {pages} {403} (\bibinfo {year} {1999})},\ \Eprint
  {http://arxiv.org/abs/hep-ph/9907312} {arXiv:hep-ph/9907312 [hep-ph]}
  \BibitemShut {NoStop}%
\bibitem [{\citenamefont {Hatta}\ and\ \citenamefont
  {Ikeda}(2003)}]{hatta2003universality}%
  \BibitemOpen
  \bibfield  {author} {\bibinfo {author} {\bibfnamefont {Y.}~\bibnamefont
  {Hatta}}\ and\ \bibinfo {author} {\bibfnamefont {T.}~\bibnamefont {Ikeda}},\
  }\href {\doibase 10.1103/PhysRevD.67.014028} {\bibfield  {journal} {\bibinfo
  {journal} {Phys. Rev.}\ }\textbf {\bibinfo {volume} {D67}},\ \bibinfo {pages}
  {014028} (\bibinfo {year} {2003})},\ \Eprint
  {http://arxiv.org/abs/hep-ph/0210284} {arXiv:hep-ph/0210284 [hep-ph]}
  \BibitemShut {NoStop}%
\bibitem [{\citenamefont {Costa}\ \emph {et~al.}(2008)\citenamefont {Costa},
  \citenamefont {Ruivo},\ and\ \citenamefont
  {de~Sousa}}]{costa2008thermodynamics}%
  \BibitemOpen
  \bibfield  {author} {\bibinfo {author} {\bibfnamefont {P.}~\bibnamefont
  {Costa}}, \bibinfo {author} {\bibfnamefont {M.~C.}\ \bibnamefont {Ruivo}}, \
  and\ \bibinfo {author} {\bibfnamefont {C.~A.}\ \bibnamefont {de~Sousa}},\
  }\href {\doibase 10.1103/PhysRevD.77.096001} {\bibfield  {journal} {\bibinfo
  {journal} {Phys. Rev.}\ }\textbf {\bibinfo {volume} {D77}},\ \bibinfo {pages}
  {096001} (\bibinfo {year} {2008})},\ \Eprint {http://arxiv.org/abs/0801.3417}
  {arXiv:0801.3417 [hep-ph]} \BibitemShut {NoStop}%
\bibitem [{\citenamefont {Sasaki}\ \emph {et~al.}(2007)\citenamefont {Sasaki},
  \citenamefont {Friman},\ and\ \citenamefont
  {Redlich}}]{sasaki2007susceptibilities}%
  \BibitemOpen
  \bibfield  {author} {\bibinfo {author} {\bibfnamefont {C.}~\bibnamefont
  {Sasaki}}, \bibinfo {author} {\bibfnamefont {B.}~\bibnamefont {Friman}}, \
  and\ \bibinfo {author} {\bibfnamefont {K.}~\bibnamefont {Redlich}},\ }\href
  {\doibase 10.1103/PhysRevD.75.074013} {\bibfield  {journal} {\bibinfo
  {journal} {Phys. Rev.}\ }\textbf {\bibinfo {volume} {D75}},\ \bibinfo {pages}
  {074013} (\bibinfo {year} {2007})},\ \Eprint
  {http://arxiv.org/abs/hep-ph/0611147} {arXiv:hep-ph/0611147 [hep-ph]}
  \BibitemShut {NoStop}%
\bibitem [{\citenamefont {Schaefer}\ and\ \citenamefont
  {Wambach}(2007)}]{schaefer2007susceptibilities}%
  \BibitemOpen
  \bibfield  {author} {\bibinfo {author} {\bibfnamefont {B.-J.}\ \bibnamefont
  {Schaefer}}\ and\ \bibinfo {author} {\bibfnamefont {J.}~\bibnamefont
  {Wambach}},\ }\href {\doibase 10.1103/PhysRevD.75.085015} {\bibfield
  {journal} {\bibinfo  {journal} {Phys. Rev.}\ }\textbf {\bibinfo {volume}
  {D75}},\ \bibinfo {pages} {085015} (\bibinfo {year} {2007})},\ \Eprint
  {http://arxiv.org/abs/hep-ph/0603256} {arXiv:hep-ph/0603256 [hep-ph]}
  \BibitemShut {NoStop}%
\bibitem [{\citenamefont {Redlich}\ \emph {et~al.}(2008)\citenamefont
  {Redlich}, \citenamefont {Friman},\ and\ \citenamefont
  {Sasaki}}]{redlich2008density}%
  \BibitemOpen
  \bibfield  {author} {\bibinfo {author} {\bibfnamefont {K.}~\bibnamefont
  {Redlich}}, \bibinfo {author} {\bibfnamefont {B.}~\bibnamefont {Friman}}, \
  and\ \bibinfo {author} {\bibfnamefont {C.}~\bibnamefont {Sasaki}},\
  }\bibfield  {booktitle} {\emph {\bibinfo {booktitle} {{Strangeness in quark
  matter. Proceedings, International Conference, SQM 2007, Levoca, Slovakia,
  June 24-29, 2007}}},\ }\href {\doibase 10.1088/0954-3899/35/4/044013}
  {\bibfield  {journal} {\bibinfo  {journal} {J. Phys.}\ }\textbf {\bibinfo
  {volume} {G35}},\ \bibinfo {pages} {044013} (\bibinfo {year} {2008})},\
  \Eprint {http://arxiv.org/abs/0712.2926} {arXiv:0712.2926 [hep-ph]}
  \BibitemShut {NoStop}%
\bibitem [{\citenamefont {Fukushima}(2008)}]{fukushima2008phase}%
  \BibitemOpen
  \bibfield  {author} {\bibinfo {author} {\bibfnamefont {K.}~\bibnamefont
  {Fukushima}},\ }\href {\doibase 10.1103/PhysRevD.77.114028,
  10.1103/PhysRevD.78.039902} {\bibfield  {journal} {\bibinfo  {journal} {Phys.
  Rev.}\ }\textbf {\bibinfo {volume} {D77}},\ \bibinfo {pages} {114028}
  (\bibinfo {year} {2008})},\ \bibinfo {note} {[Erratum: Phys.
  Rev.D78,039902(2008)]},\ \Eprint {http://arxiv.org/abs/0803.3318}
  {arXiv:0803.3318 [hep-ph]} \BibitemShut {NoStop}%
\bibitem [{\citenamefont {Cui}\ \emph {et~al.}(2013)\citenamefont {Cui},
  \citenamefont {Shi}, \citenamefont {Xia}, \citenamefont {Jiang},\ and\
  \citenamefont {Zong}}]{cui2013wigner}%
  \BibitemOpen
  \bibfield  {author} {\bibinfo {author} {\bibfnamefont {Z.-F.}\ \bibnamefont
  {Cui}}, \bibinfo {author} {\bibfnamefont {C.}~\bibnamefont {Shi}}, \bibinfo
  {author} {\bibfnamefont {Y.-H.}\ \bibnamefont {Xia}}, \bibinfo {author}
  {\bibfnamefont {Y.}~\bibnamefont {Jiang}}, \ and\ \bibinfo {author}
  {\bibfnamefont {H.-S.}\ \bibnamefont {Zong}},\ }\href {\doibase
  10.1140/epjc/s10052-013-2612-6} {\bibfield  {journal} {\bibinfo  {journal}
  {Eur. Phys. J.}\ }\textbf {\bibinfo {volume} {C73}},\ \bibinfo {pages} {2612}
  (\bibinfo {year} {2013})}\BibitemShut {NoStop}%
\bibitem [{\citenamefont {Jiang}\ \emph {et~al.}(2013)\citenamefont {Jiang},
  \citenamefont {Chen}, \citenamefont {Sun},\ and\ \citenamefont
  {Zong}}]{jiang2013chiral}%
  \BibitemOpen
  \bibfield  {author} {\bibinfo {author} {\bibfnamefont {Y.}~\bibnamefont
  {Jiang}}, \bibinfo {author} {\bibfnamefont {H.}~\bibnamefont {Chen}},
  \bibinfo {author} {\bibfnamefont {W.-M.}\ \bibnamefont {Sun}}, \ and\
  \bibinfo {author} {\bibfnamefont {H.-S.}\ \bibnamefont {Zong}},\ }\href
  {\doibase 10.1007/JHEP04(2013)014} {\bibfield  {journal} {\bibinfo  {journal}
  {JHEP}\ }\textbf {\bibinfo {volume} {04}},\ \bibinfo {pages} {014} (\bibinfo
  {year} {2013})}\BibitemShut {NoStop}%
\bibitem [{\citenamefont {Cui}\ \emph {et~al.}(2014)\citenamefont {Cui},
  \citenamefont {Shi}, \citenamefont {Sun}, \citenamefont {Wang},\ and\
  \citenamefont {Zong}}]{Cui2013aba}%
  \BibitemOpen
  \bibfield  {author} {\bibinfo {author} {\bibfnamefont {Z.-f.}\ \bibnamefont
  {Cui}}, \bibinfo {author} {\bibfnamefont {C.}~\bibnamefont {Shi}}, \bibinfo
  {author} {\bibfnamefont {W.-m.}\ \bibnamefont {Sun}}, \bibinfo {author}
  {\bibfnamefont {Y.-l.}\ \bibnamefont {Wang}}, \ and\ \bibinfo {author}
  {\bibfnamefont {H.-s.}\ \bibnamefont {Zong}},\ }\href {\doibase
  10.1140/epjc/s10052-014-2782-x} {\bibfield  {journal} {\bibinfo  {journal}
  {Eur. Phys. J.}\ }\textbf {\bibinfo {volume} {C74}},\ \bibinfo {pages} {2782}
  (\bibinfo {year} {2014})},\ \Eprint {http://arxiv.org/abs/1311.4014}
  {arXiv:1311.4014 [hep-ph]} \BibitemShut {NoStop}%
\bibitem [{\citenamefont {Wang}\ \emph {et~al.}(2015)\citenamefont {Wang},
  \citenamefont {Wang}, \citenamefont {Cui},\ and\ \citenamefont
  {Zong}}]{Wang2015tia}%
  \BibitemOpen
  \bibfield  {author} {\bibinfo {author} {\bibfnamefont {B.}~\bibnamefont
  {Wang}}, \bibinfo {author} {\bibfnamefont {Y.-L.}\ \bibnamefont {Wang}},
  \bibinfo {author} {\bibfnamefont {Z.-F.}\ \bibnamefont {Cui}}, \ and\
  \bibinfo {author} {\bibfnamefont {H.-S.}\ \bibnamefont {Zong}},\ }\href
  {\doibase 10.1103/PhysRevD.91.034017} {\bibfield  {journal} {\bibinfo
  {journal} {Phys. Rev.}\ }\textbf {\bibinfo {volume} {D91}},\ \bibinfo {pages}
  {034017} (\bibinfo {year} {2015})}\BibitemShut {NoStop}%
\bibitem [{\citenamefont {Shi}\ \emph {et~al.}(2015)\citenamefont {Shi},
  \citenamefont {Yang}, \citenamefont {Xia}, \citenamefont {Cui}, \citenamefont
  {Liu},\ and\ \citenamefont {Zong}}]{Shi2015ufa}%
  \BibitemOpen
  \bibfield  {author} {\bibinfo {author} {\bibfnamefont {S.}~\bibnamefont
  {Shi}}, \bibinfo {author} {\bibfnamefont {Y.-C.}\ \bibnamefont {Yang}},
  \bibinfo {author} {\bibfnamefont {Y.-H.}\ \bibnamefont {Xia}}, \bibinfo
  {author} {\bibfnamefont {Z.-F.}\ \bibnamefont {Cui}}, \bibinfo {author}
  {\bibfnamefont {X.-J.}\ \bibnamefont {Liu}}, \ and\ \bibinfo {author}
  {\bibfnamefont {H.-S.}\ \bibnamefont {Zong}},\ }\href {\doibase
  10.1103/PhysRevD.91.036006} {\bibfield  {journal} {\bibinfo  {journal} {Phys.
  Rev.}\ }\textbf {\bibinfo {volume} {D91}},\ \bibinfo {pages} {036006}
  (\bibinfo {year} {2015})},\ \Eprint {http://arxiv.org/abs/1503.00452}
  {arXiv:1503.00452 [hep-ph]} \BibitemShut {NoStop}%
\bibitem [{\citenamefont {Cui}\ \emph {et~al.}(2015)\citenamefont {Cui},
  \citenamefont {Hou}, \citenamefont {Shi}, \citenamefont {Wang},\ and\
  \citenamefont {Zong}}]{Cui2015xta}%
  \BibitemOpen
  \bibfield  {author} {\bibinfo {author} {\bibfnamefont {Z.-F.}\ \bibnamefont
  {Cui}}, \bibinfo {author} {\bibfnamefont {F.-Y.}\ \bibnamefont {Hou}},
  \bibinfo {author} {\bibfnamefont {Y.-M.}\ \bibnamefont {Shi}}, \bibinfo
  {author} {\bibfnamefont {Y.-L.}\ \bibnamefont {Wang}}, \ and\ \bibinfo
  {author} {\bibfnamefont {H.-S.}\ \bibnamefont {Zong}},\ }\href {\doibase
  10.1016/j.aop.2015.03.025} {\bibfield  {journal} {\bibinfo  {journal} {Annals
  Phys.}\ }\textbf {\bibinfo {volume} {358}},\ \bibinfo {pages} {172} (\bibinfo
  {year} {2015})},\ \Eprint {http://arxiv.org/abs/1505.00310} {arXiv:1505.00310
  [hep-ph]} \BibitemShut {NoStop}%
\bibitem [{\citenamefont {Cui}\ \emph {et~al.}(2017)\citenamefont {Cui},
  \citenamefont {Zhang},\ and\ \citenamefont {Zong}}]{Cui2017ilj}%
  \BibitemOpen
  \bibfield  {author} {\bibinfo {author} {\bibfnamefont {Z.-F.}\ \bibnamefont
  {Cui}}, \bibinfo {author} {\bibfnamefont {J.-L.}\ \bibnamefont {Zhang}}, \
  and\ \bibinfo {author} {\bibfnamefont {H.-S.}\ \bibnamefont {Zong}},\ }\href
  {\doibase 10.1038/srep45937} {\bibfield  {journal} {\bibinfo  {journal} {Sci.
  Rep.}\ }\textbf {\bibinfo {volume} {7}},\ \bibinfo {pages} {45937} (\bibinfo
  {year} {2017})}\BibitemShut {NoStop}%
\bibitem [{\citenamefont {Allton}\ \emph {et~al.}(2005)\citenamefont {Allton},
  \citenamefont {Doring}, \citenamefont {Ejiri}, \citenamefont {Hands},
  \citenamefont {Kaczmarek}, \citenamefont {Karsch}, \citenamefont {Laermann},\
  and\ \citenamefont {Redlich}}]{allton2005thermodynamics}%
  \BibitemOpen
  \bibfield  {author} {\bibinfo {author} {\bibfnamefont {C.~R.}\ \bibnamefont
  {Allton}}, \bibinfo {author} {\bibfnamefont {M.}~\bibnamefont {Doring}},
  \bibinfo {author} {\bibfnamefont {S.}~\bibnamefont {Ejiri}}, \bibinfo
  {author} {\bibfnamefont {S.~J.}\ \bibnamefont {Hands}}, \bibinfo {author}
  {\bibfnamefont {O.}~\bibnamefont {Kaczmarek}}, \bibinfo {author}
  {\bibfnamefont {F.}~\bibnamefont {Karsch}}, \bibinfo {author} {\bibfnamefont
  {E.}~\bibnamefont {Laermann}}, \ and\ \bibinfo {author} {\bibfnamefont
  {K.}~\bibnamefont {Redlich}},\ }\href {\doibase 10.1103/PhysRevD.71.054508}
  {\bibfield  {journal} {\bibinfo  {journal} {Phys. Rev.}\ }\textbf {\bibinfo
  {volume} {D71}},\ \bibinfo {pages} {054508} (\bibinfo {year} {2005})},\
  \Eprint {http://arxiv.org/abs/hep-lat/0501030} {arXiv:hep-lat/0501030
  [hep-lat]} \BibitemShut {NoStop}%
\bibitem [{\citenamefont {Gavai}\ and\ \citenamefont
  {Gupta}(2005)}]{gavai2005critical}%
  \BibitemOpen
  \bibfield  {author} {\bibinfo {author} {\bibfnamefont {R.~V.}\ \bibnamefont
  {Gavai}}\ and\ \bibinfo {author} {\bibfnamefont {S.}~\bibnamefont {Gupta}},\
  }\href {\doibase 10.1103/PhysRevD.71.114014} {\bibfield  {journal} {\bibinfo
  {journal} {Phys. Rev.}\ }\textbf {\bibinfo {volume} {D71}},\ \bibinfo {pages}
  {114014} (\bibinfo {year} {2005})},\ \Eprint
  {http://arxiv.org/abs/hep-lat/0412035} {arXiv:hep-lat/0412035 [hep-lat]}
  \BibitemShut {NoStop}%
\bibitem [{\citenamefont {Philipsen}(2007)}]{philipsen2007lattice}%
  \BibitemOpen
  \bibfield  {author} {\bibinfo {author} {\bibfnamefont {O.}~\bibnamefont
  {Philipsen}},\ }\bibfield  {booktitle} {\emph {\bibinfo {booktitle}
  {{Conceptual and Numerical Challenges in Femto- and Peta-Scale Physics.
  Proceedings, 45. Internationale Universitätswochen für theoretische Physik:
  Schladming, Austria, February 24-March 3, 2007}}},\ }\href {\doibase
  10.1140/epjst/e2007-00376-3} {\bibfield  {journal} {\bibinfo  {journal} {Eur.
  Phys. J. ST}\ }\textbf {\bibinfo {volume} {152}},\ \bibinfo {pages} {29}
  (\bibinfo {year} {2007})},\ \Eprint {http://arxiv.org/abs/0708.1293}
  {arXiv:0708.1293 [hep-lat]} \BibitemShut {NoStop}%
\bibitem [{\citenamefont {Aggarwal}\ \emph {et~al.}(2010)\citenamefont
  {Aggarwal} \emph {et~al.}}]{aggarwal2010higher}%
  \BibitemOpen
  \bibfield  {author} {\bibinfo {author} {\bibfnamefont {M.~M.}\ \bibnamefont
  {Aggarwal}} \emph {et~al.} (\bibinfo {collaboration} {STAR}),\ }\href
  {\doibase 10.1103/PhysRevLett.105.022302} {\bibfield  {journal} {\bibinfo
  {journal} {Phys. Rev. Lett.}\ }\textbf {\bibinfo {volume} {105}},\ \bibinfo
  {pages} {022302} (\bibinfo {year} {2010})},\ \Eprint
  {http://arxiv.org/abs/1004.4959} {arXiv:1004.4959 [nucl-ex]} \BibitemShut
  {NoStop}%
\bibitem [{\citenamefont {Gupta}\ \emph {et~al.}(2011)\citenamefont {Gupta},
  \citenamefont {Luo}, \citenamefont {Mohanty}, \citenamefont {Ritter},\ and\
  \citenamefont {Xu}}]{gupta2011scale}%
  \BibitemOpen
  \bibfield  {author} {\bibinfo {author} {\bibfnamefont {S.}~\bibnamefont
  {Gupta}}, \bibinfo {author} {\bibfnamefont {X.}~\bibnamefont {Luo}}, \bibinfo
  {author} {\bibfnamefont {B.}~\bibnamefont {Mohanty}}, \bibinfo {author}
  {\bibfnamefont {H.~G.}\ \bibnamefont {Ritter}}, \ and\ \bibinfo {author}
  {\bibfnamefont {N.}~\bibnamefont {Xu}},\ }\href {\doibase
  10.1126/science.1204621} {\bibfield  {journal} {\bibinfo  {journal}
  {Science}\ }\textbf {\bibinfo {volume} {332}},\ \bibinfo {pages} {1525}
  (\bibinfo {year} {2011})},\ \Eprint {http://arxiv.org/abs/1105.3934}
  {arXiv:1105.3934 [hep-ph]} \BibitemShut {NoStop}%
\bibitem [{\citenamefont {Jeon}\ and\ \citenamefont
  {Koch}(2000)}]{jeon2000charged}%
  \BibitemOpen
  \bibfield  {author} {\bibinfo {author} {\bibfnamefont {S.}~\bibnamefont
  {Jeon}}\ and\ \bibinfo {author} {\bibfnamefont {V.}~\bibnamefont {Koch}},\
  }\href {\doibase 10.1103/PhysRevLett.85.2076} {\bibfield  {journal} {\bibinfo
   {journal} {Phys. Rev. Lett.}\ }\textbf {\bibinfo {volume} {85}},\ \bibinfo
  {pages} {2076} (\bibinfo {year} {2000})},\ \Eprint
  {http://arxiv.org/abs/hep-ph/0003168} {arXiv:hep-ph/0003168 [hep-ph]}
  \BibitemShut {NoStop}%
\bibitem [{\citenamefont {Zong}\ \emph {et~al.}(2008)\citenamefont {Zong},
  \citenamefont {Hou}, \citenamefont {Sun},\ and\ \citenamefont
  {Lu}}]{zong2008calculation}%
  \BibitemOpen
  \bibfield  {author} {\bibinfo {author} {\bibfnamefont {H.-S.}\ \bibnamefont
  {Zong}}, \bibinfo {author} {\bibfnamefont {F.-Y.}\ \bibnamefont {Hou}},
  \bibinfo {author} {\bibfnamefont {W.-M.}\ \bibnamefont {Sun}}, \ and\
  \bibinfo {author} {\bibfnamefont {X.-F.}\ \bibnamefont {Lu}},\ }\href
  {\doibase 10.1088/0253-6102/49/5/40} {\bibfield  {journal} {\bibinfo
  {journal} {Commun. Theor. Phys.}\ }\textbf {\bibinfo {volume} {49}},\
  \bibinfo {pages} {1269} (\bibinfo {year} {2008})}\BibitemShut {NoStop}%
\bibitem [{\citenamefont {Roessner}\ \emph
  {et~al.}(2007{\natexlab{a}})\citenamefont {Roessner}, \citenamefont {Ratti},\
  and\ \citenamefont {Weise}}]{R2006Polyakov}%
  \BibitemOpen
  \bibfield  {author} {\bibinfo {author} {\bibfnamefont {S.}~\bibnamefont
  {Roessner}}, \bibinfo {author} {\bibfnamefont {C.}~\bibnamefont {Ratti}}, \
  and\ \bibinfo {author} {\bibfnamefont {W.}~\bibnamefont {Weise}},\ }\href
  {\doibase 10.1103/PhysRevD.75.034007} {\bibfield  {journal} {\bibinfo
  {journal} {Phys. Rev.}\ }\textbf {\bibinfo {volume} {D75}},\ \bibinfo {pages}
  {034007} (\bibinfo {year} {2007}{\natexlab{a}})},\ \Eprint
  {http://arxiv.org/abs/hep-ph/0609281} {arXiv:hep-ph/0609281 [hep-ph]}
  \BibitemShut {NoStop}%
\bibitem [{\citenamefont {Ruggieri}(2011{\natexlab{a}})}]{Ruggieri2011The}%
  \BibitemOpen
  \bibfield  {author} {\bibinfo {author} {\bibfnamefont {M.}~\bibnamefont
  {Ruggieri}},\ }\href {\doibase 10.1103/PhysRevD.84.014011} {\bibfield
  {journal} {\bibinfo  {journal} {Phys. Rev.}\ }\textbf {\bibinfo {volume}
  {D84}},\ \bibinfo {pages} {014011} (\bibinfo {year} {2011}{\natexlab{a}})},\
  \Eprint {http://arxiv.org/abs/1103.6186} {arXiv:1103.6186 [hep-ph]}
  \BibitemShut {NoStop}%
\bibitem [{\citenamefont {Fukushima}(2004)}]{fukushima2004chiral}%
  \BibitemOpen
  \bibfield  {author} {\bibinfo {author} {\bibfnamefont {K.}~\bibnamefont
  {Fukushima}},\ }\href {\doibase 10.1016/j.physletb.2004.04.027} {\bibfield
  {journal} {\bibinfo  {journal} {Phys. Lett.}\ }\textbf {\bibinfo {volume}
  {B591}},\ \bibinfo {pages} {277} (\bibinfo {year} {2004})},\ \Eprint
  {http://arxiv.org/abs/hep-ph/0310121} {arXiv:hep-ph/0310121 [hep-ph]}
  \BibitemShut {NoStop}%
\bibitem [{\citenamefont {Zong}\ and\ \citenamefont {Sun}(2008)}]{Zong2008sm}%
  \BibitemOpen
  \bibfield  {author} {\bibinfo {author} {\bibfnamefont {H.-s.}\ \bibnamefont
  {Zong}}\ and\ \bibinfo {author} {\bibfnamefont {W.-m.}\ \bibnamefont {Sun}},\
  }\href {\doibase 10.1103/PhysRevD.78.054001} {\bibfield  {journal} {\bibinfo
  {journal} {Phys. Rev.}\ }\textbf {\bibinfo {volume} {D78}},\ \bibinfo {pages}
  {054001} (\bibinfo {year} {2008})},\ \Eprint {http://arxiv.org/abs/0810.2843}
  {arXiv:0810.2843 [hep-ph]} \BibitemShut {NoStop}%
\bibitem [{\citenamefont {Hou}\ \emph {et~al.}(2005)\citenamefont {Hou},
  \citenamefont {Chang}, \citenamefont {Sun}, \citenamefont {Zong},\ and\
  \citenamefont {Liu}}]{hou2005new}%
  \BibitemOpen
  \bibfield  {author} {\bibinfo {author} {\bibfnamefont {F.-y.}\ \bibnamefont
  {Hou}}, \bibinfo {author} {\bibfnamefont {L.}~\bibnamefont {Chang}}, \bibinfo
  {author} {\bibfnamefont {W.-m.}\ \bibnamefont {Sun}}, \bibinfo {author}
  {\bibfnamefont {H.-s.}\ \bibnamefont {Zong}}, \ and\ \bibinfo {author}
  {\bibfnamefont {Y.-x.}\ \bibnamefont {Liu}},\ }\href {\doibase
  10.1103/PhysRevC.72.034901} {\bibfield  {journal} {\bibinfo  {journal} {Phys.
  Rev.}\ }\textbf {\bibinfo {volume} {C72}},\ \bibinfo {pages} {034901}
  (\bibinfo {year} {2005})},\ \Eprint {http://arxiv.org/abs/hep-ph/0504281}
  {arXiv:hep-ph/0504281 [hep-ph]} \BibitemShut {NoStop}%
\bibitem [{\citenamefont {Gavai}\ and\ \citenamefont
  {Gupta}(2003)}]{gavai2003pressure}%
  \BibitemOpen
  \bibfield  {author} {\bibinfo {author} {\bibfnamefont {R.~V.}\ \bibnamefont
  {Gavai}}\ and\ \bibinfo {author} {\bibfnamefont {S.}~\bibnamefont {Gupta}},\
  }\href {\doibase 10.1103/PhysRevD.68.034506} {\bibfield  {journal} {\bibinfo
  {journal} {Phys. Rev.}\ }\textbf {\bibinfo {volume} {D68}},\ \bibinfo {pages}
  {034506} (\bibinfo {year} {2003})},\ \Eprint
  {http://arxiv.org/abs/hep-lat/0303013} {arXiv:hep-lat/0303013 [hep-lat]}
  \BibitemShut {NoStop}%
\bibitem [{\citenamefont {Ratti}\ \emph {et~al.}(2007)\citenamefont {Ratti},
  \citenamefont {Roessner}, \citenamefont {Thaler},\ and\ \citenamefont
  {Weise}}]{Ratti2006wg}%
  \BibitemOpen
  \bibfield  {author} {\bibinfo {author} {\bibfnamefont {C.}~\bibnamefont
  {Ratti}}, \bibinfo {author} {\bibfnamefont {S.}~\bibnamefont {Roessner}},
  \bibinfo {author} {\bibfnamefont {M.~A.}\ \bibnamefont {Thaler}}, \ and\
  \bibinfo {author} {\bibfnamefont {W.}~\bibnamefont {Weise}},\ }\bibfield
  {booktitle} {\emph {\bibinfo {booktitle} {{Proceedings, Workshop for Young
  Scientists on the Physics of Ultrarelativistic Nucleus-Nucleus Collisions
  (Hot Quarks 2006): Villasimius, Italy, May 15-20, 2006}}},\ }\href {\doibase
  10.1140/epjc/s10052-006-0065-x} {\bibfield  {journal} {\bibinfo  {journal}
  {Eur. Phys. J.}\ }\textbf {\bibinfo {volume} {C49}},\ \bibinfo {pages} {213}
  (\bibinfo {year} {2007})},\ \Eprint {http://arxiv.org/abs/hep-ph/0609218}
  {arXiv:hep-ph/0609218 [hep-ph]} \BibitemShut {NoStop}%
\bibitem [{\citenamefont {Cui}\ \emph {et~al.}(2016)\citenamefont {Cui},
  \citenamefont {Cloet}, \citenamefont {Lu}, \citenamefont {Roberts},
  \citenamefont {Schmidt}, \citenamefont {Xu},\ and\ \citenamefont
  {Zong}}]{cui2016critical}%
  \BibitemOpen
  \bibfield  {author} {\bibinfo {author} {\bibfnamefont {Z.-F.}\ \bibnamefont
  {Cui}}, \bibinfo {author} {\bibfnamefont {I.~C.}\ \bibnamefont {Cloet}},
  \bibinfo {author} {\bibfnamefont {Y.}~\bibnamefont {Lu}}, \bibinfo {author}
  {\bibfnamefont {C.~D.}\ \bibnamefont {Roberts}}, \bibinfo {author}
  {\bibfnamefont {S.~M.}\ \bibnamefont {Schmidt}}, \bibinfo {author}
  {\bibfnamefont {S.-S.}\ \bibnamefont {Xu}}, \ and\ \bibinfo {author}
  {\bibfnamefont {H.-S.}\ \bibnamefont {Zong}},\ }\href {\doibase
  10.1103/PhysRevD.94.071503} {\bibfield  {journal} {\bibinfo  {journal} {Phys.
  Rev.}\ }\textbf {\bibinfo {volume} {D94}},\ \bibinfo {pages} {071503}
  (\bibinfo {year} {2016})},\ \Eprint {http://arxiv.org/abs/1604.08454}
  {arXiv:1604.08454 [nucl-th]} \BibitemShut {NoStop}%
\bibitem [{\citenamefont {Pan}\ \emph {et~al.}(2017)\citenamefont {Pan},
  \citenamefont {Cui}, \citenamefont {Chang},\ and\ \citenamefont
  {Zong}}]{Pan2016ecs}%
  \BibitemOpen
  \bibfield  {author} {\bibinfo {author} {\bibfnamefont {Z.}~\bibnamefont
  {Pan}}, \bibinfo {author} {\bibfnamefont {Z.-F.}\ \bibnamefont {Cui}},
  \bibinfo {author} {\bibfnamefont {C.-H.}\ \bibnamefont {Chang}}, \ and\
  \bibinfo {author} {\bibfnamefont {H.-S.}\ \bibnamefont {Zong}},\ }\href
  {\doibase 10.1142/S0217751X17500671} {\bibfield  {journal} {\bibinfo
  {journal} {Int. J. Mod. Phys.}\ }\textbf {\bibinfo {volume} {A32}},\ \bibinfo
  {pages} {1750067} (\bibinfo {year} {2017})},\ \Eprint
  {http://arxiv.org/abs/1611.07370} {arXiv:1611.07370 [hep-ph]} \BibitemShut
  {NoStop}%
\bibitem [{\citenamefont {Roessner}\ \emph
  {et~al.}(2007{\natexlab{b}})\citenamefont {Roessner}, \citenamefont {Ratti},\
  and\ \citenamefont {Weise}}]{roessner2007polyakov}%
  \BibitemOpen
  \bibfield  {author} {\bibinfo {author} {\bibfnamefont {S.}~\bibnamefont
  {Roessner}}, \bibinfo {author} {\bibfnamefont {C.}~\bibnamefont {Ratti}}, \
  and\ \bibinfo {author} {\bibfnamefont {W.}~\bibnamefont {Weise}},\ }\href
  {\doibase 10.1103/PhysRevD.75.034007} {\bibfield  {journal} {\bibinfo
  {journal} {Phys. Rev.}\ }\textbf {\bibinfo {volume} {D75}},\ \bibinfo {pages}
  {034007} (\bibinfo {year} {2007}{\natexlab{b}})},\ \Eprint
  {http://arxiv.org/abs/hep-ph/0609281} {arXiv:hep-ph/0609281 [hep-ph]}
  \BibitemShut {NoStop}%
\bibitem [{\citenamefont {Roessner}\ \emph {et~al.}(2008)\citenamefont
  {Roessner}, \citenamefont {Hell}, \citenamefont {Ratti},\ and\ \citenamefont
  {Weise}}]{Rossner2007ik}%
  \BibitemOpen
  \bibfield  {author} {\bibinfo {author} {\bibfnamefont {S.}~\bibnamefont
  {Roessner}}, \bibinfo {author} {\bibfnamefont {T.}~\bibnamefont {Hell}},
  \bibinfo {author} {\bibfnamefont {C.}~\bibnamefont {Ratti}}, \ and\ \bibinfo
  {author} {\bibfnamefont {W.}~\bibnamefont {Weise}},\ }\href {\doibase
  10.1016/j.nuclphysa.2008.10.006} {\bibfield  {journal} {\bibinfo  {journal}
  {Nucl. Phys.}\ }\textbf {\bibinfo {volume} {A814}},\ \bibinfo {pages} {118}
  (\bibinfo {year} {2008})},\ \Eprint {http://arxiv.org/abs/0712.3152}
  {arXiv:0712.3152 [hep-ph]} \BibitemShut {NoStop}%
\bibitem [{\citenamefont {Ruggieri}(2011{\natexlab{b}})}]{Ruggieri:2011xc}%
  \BibitemOpen
  \bibfield  {author} {\bibinfo {author} {\bibfnamefont {M.}~\bibnamefont
  {Ruggieri}},\ }\href {\doibase 10.1103/PhysRevD.84.014011} {\bibfield
  {journal} {\bibinfo  {journal} {Phys. Rev.}\ }\textbf {\bibinfo {volume}
  {D84}},\ \bibinfo {pages} {014011} (\bibinfo {year} {2011}{\natexlab{b}})},\
  \Eprint {http://arxiv.org/abs/1103.6186} {arXiv:1103.6186 [hep-ph]}
  \BibitemShut {NoStop}%
\bibitem [{\citenamefont {Cleymans}\ and\ \citenamefont
  {Redlich}(1998)}]{cleymans1998unified}%
  \BibitemOpen
  \bibfield  {author} {\bibinfo {author} {\bibfnamefont {J.}~\bibnamefont
  {Cleymans}}\ and\ \bibinfo {author} {\bibfnamefont {K.}~\bibnamefont
  {Redlich}},\ }\href {\doibase 10.1103/PhysRevLett.81.5284} {\bibfield
  {journal} {\bibinfo  {journal} {Phys. Rev. Lett.}\ }\textbf {\bibinfo
  {volume} {81}},\ \bibinfo {pages} {5284} (\bibinfo {year} {1998})},\ \Eprint
  {http://arxiv.org/abs/nucl-th/9808030} {arXiv:nucl-th/9808030 [nucl-th]}
  \BibitemShut {NoStop}%
\bibitem [{\citenamefont {Braun-Munzinger}\ and\ \citenamefont
  {Wambach}(2009)}]{braun2009colloquium}%
  \BibitemOpen
  \bibfield  {author} {\bibinfo {author} {\bibfnamefont {P.}~\bibnamefont
  {Braun-Munzinger}}\ and\ \bibinfo {author} {\bibfnamefont {J.}~\bibnamefont
  {Wambach}},\ }\href {\doibase 10.1103/RevModPhys.81.1031} {\bibfield
  {journal} {\bibinfo  {journal} {Rev. Mod. Phys.}\ }\textbf {\bibinfo {volume}
  {81}},\ \bibinfo {pages} {1031} (\bibinfo {year} {2009})},\ \Eprint
  {http://arxiv.org/abs/0801.4256} {arXiv:0801.4256 [hep-ph]} \BibitemShut
  {NoStop}%
\bibitem [{\citenamefont {Cleymans}\ \emph {et~al.}(2006)\citenamefont
  {Cleymans}, \citenamefont {Oeschler}, \citenamefont {Redlich},\ and\
  \citenamefont {Wheaton}}]{cleymans2006comparison}%
  \BibitemOpen
  \bibfield  {author} {\bibinfo {author} {\bibfnamefont {J.}~\bibnamefont
  {Cleymans}}, \bibinfo {author} {\bibfnamefont {H.}~\bibnamefont {Oeschler}},
  \bibinfo {author} {\bibfnamefont {K.}~\bibnamefont {Redlich}}, \ and\
  \bibinfo {author} {\bibfnamefont {S.}~\bibnamefont {Wheaton}},\ }\href
  {\doibase 10.1103/PhysRevC.73.034905} {\bibfield  {journal} {\bibinfo
  {journal} {Phys. Rev.}\ }\textbf {\bibinfo {volume} {C73}},\ \bibinfo {pages}
  {034905} (\bibinfo {year} {2006})},\ \Eprint
  {http://arxiv.org/abs/hep-ph/0511094} {arXiv:hep-ph/0511094 [hep-ph]}
  \BibitemShut {NoStop}%
\bibitem [{\citenamefont {Andronic}\ and\ \citenamefont
  {Braun-Munzinger}(2004)}]{andronic2004ultrarelativistic}%
  \BibitemOpen
  \bibfield  {author} {\bibinfo {author} {\bibfnamefont {A.}~\bibnamefont
  {Andronic}}\ and\ \bibinfo {author} {\bibfnamefont {P.}~\bibnamefont
  {Braun-Munzinger}},\ }\bibfield  {booktitle} {\emph {\bibinfo {booktitle}
  {{Proceedings, 8th Hispalensis International Summer School on Exotic Nuclear
  Physics: The Hispalensis Lectures on Nuclear Physics: Seville, Spain, June
  9-21, 2003}}},\ }\href {\doibase 10.1007/978-3-540-44504-3_2} {\bibfield
  {journal} {\bibinfo  {journal} {Lect. Notes Phys.}\ }\textbf {\bibinfo
  {volume} {652}},\ \bibinfo {pages} {35} (\bibinfo {year} {2004})},\ \Eprint
  {http://arxiv.org/abs/hep-ph/0402291} {arXiv:hep-ph/0402291 [hep-ph]}
  \BibitemShut {NoStop}%
\bibitem [{\citenamefont {Gavai}\ and\ \citenamefont
  {Gupta}(2011)}]{gavai2011lattice}%
  \BibitemOpen
  \bibfield  {author} {\bibinfo {author} {\bibfnamefont {R.~V.}\ \bibnamefont
  {Gavai}}\ and\ \bibinfo {author} {\bibfnamefont {S.}~\bibnamefont {Gupta}},\
  }\href {\doibase 10.1016/j.physletb.2011.01.006} {\bibfield  {journal}
  {\bibinfo  {journal} {Phys. Lett.}\ }\textbf {\bibinfo {volume} {B696}},\
  \bibinfo {pages} {459} (\bibinfo {year} {2011})},\ \Eprint
  {http://arxiv.org/abs/1001.3796} {arXiv:1001.3796 [hep-lat]} \BibitemShut
  {NoStop}%
\bibitem [{\citenamefont {Zhao}\ \emph {et~al.}(2014)\citenamefont {Zhao},
  \citenamefont {Cui}, \citenamefont {Jiang},\ and\ \citenamefont
  {Zong}}]{zhao2014nonlinear}%
  \BibitemOpen
  \bibfield  {author} {\bibinfo {author} {\bibfnamefont {A.-M.}\ \bibnamefont
  {Zhao}}, \bibinfo {author} {\bibfnamefont {Z.-F.}\ \bibnamefont {Cui}},
  \bibinfo {author} {\bibfnamefont {Y.}~\bibnamefont {Jiang}}, \ and\ \bibinfo
  {author} {\bibfnamefont {H.-S.}\ \bibnamefont {Zong}},\ }\href {\doibase
  10.1103/PhysRevD.90.114031} {\bibfield  {journal} {\bibinfo  {journal} {Phys.
  Rev.}\ }\textbf {\bibinfo {volume} {D90}},\ \bibinfo {pages} {114031}
  (\bibinfo {year} {2014})},\ \Eprint {http://arxiv.org/abs/1412.6884}
  {arXiv:1412.6884 [hep-ph]} \BibitemShut {NoStop}%
\bibitem [{\citenamefont {Lu}\ \emph {et~al.}(2016)\citenamefont {Lu},
  \citenamefont {Cui}, \citenamefont {Pan}, \citenamefont {Chang},\ and\
  \citenamefont {Zong}}]{Lu2016uwy}%
  \BibitemOpen
  \bibfield  {author} {\bibinfo {author} {\bibfnamefont {Y.}~\bibnamefont
  {Lu}}, \bibinfo {author} {\bibfnamefont {Z.-F.}\ \bibnamefont {Cui}},
  \bibinfo {author} {\bibfnamefont {Z.}~\bibnamefont {Pan}}, \bibinfo {author}
  {\bibfnamefont {C.-H.}\ \bibnamefont {Chang}}, \ and\ \bibinfo {author}
  {\bibfnamefont {H.-S.}\ \bibnamefont {Zong}},\ }\href {\doibase
  10.1103/PhysRevD.93.074037} {\bibfield  {journal} {\bibinfo  {journal} {Phys.
  Rev.}\ }\textbf {\bibinfo {volume} {D93}},\ \bibinfo {pages} {074037}
  (\bibinfo {year} {2016})}\BibitemShut {NoStop}%
\bibitem [{\citenamefont {Ivanytskyi}\ \emph {et~al.}(2019)\citenamefont
  {Ivanytskyi}, \citenamefont {Perez-Garcia}, \citenamefont {Sagun},\ and\
  \citenamefont {Albertus}}]{Ivanytskyi2019ojt}%
  \BibitemOpen
  \bibfield  {author} {\bibinfo {author} {\bibfnamefont {O.}~\bibnamefont
  {Ivanytskyi}}, \bibinfo {author} {\bibfnamefont {M.~A.}\ \bibnamefont
  {Perez-Garcia}}, \bibinfo {author} {\bibfnamefont {V.}~\bibnamefont {Sagun}},
  \ and\ \bibinfo {author} {\bibfnamefont {C.}~\bibnamefont {Albertus}},\
  }\href {\doibase 10.1103/PhysRevD.100.103020} {\bibfield  {journal} {\bibinfo
   {journal} {Phys. Rev.}\ }\textbf {\bibinfo {volume} {D100}},\ \bibinfo
  {pages} {103020} (\bibinfo {year} {2019})},\ \Eprint
  {http://arxiv.org/abs/1909.07421} {arXiv:1909.07421 [hep-ph]} \BibitemShut
  {NoStop}%
\bibitem [{\citenamefont {Buballa}(2005)}]{Buballa2003qv}%
  \BibitemOpen
  \bibfield  {author} {\bibinfo {author} {\bibfnamefont {M.}~\bibnamefont
  {Buballa}},\ }\href {\doibase 10.1016/j.physrep.2004.11.004} {\bibfield
  {journal} {\bibinfo  {journal} {Phys. Rept.}\ }\textbf {\bibinfo {volume}
  {407}},\ \bibinfo {pages} {205} (\bibinfo {year} {2005})},\ \Eprint
  {http://arxiv.org/abs/hep-ph/0402234} {arXiv:hep-ph/0402234 [hep-ph]}
  \BibitemShut {NoStop}%
\end{thebibliography}%


%
\end{document}